\documentclass[hyper]{JHEP3}
\pdfoutput=1

\usepackage{epsfig}
\usepackage{latexsym}
\usepackage{amsfonts}
\usepackage{amsmath}
\usepackage{amsthm}
\usepackage{amssymb}
\usepackage{amsbsy}
\usepackage{multirow}

\def\calo         {{\cal O}}

\newsavebox{\uuunit}
\sbox{\uuunit}
    {\setlength{\unitlength}{0.825em}
     \begin{picture}(0.6,0.7)
        \thinlines
        \put(0,0){\line(1,0){0.5}}
        \put(0.15,0){\line(0,1){0.7}}
        \put(0.35,0){\line(0,1){0.8}}
       \multiput(0.3,0.8)(-0.04,-0.02){12}{\rule{0.5pt}{0.5pt}}
     \end {picture}}

\def\be{\begin{equation}}
\def\ee{\end{equation}}
\def\bea{\begin{eqnarray}}
\def\eea{\end{eqnarray}}

\newcommand{\beq}{\begin{eqnarray}}
\newcommand{\eeq}{\end{eqnarray}}

\def\e{\epsilon}
\def\h{\eta}

\def\l{\lambda}
\def\L{\Lambda}
\def\k{\kappa}
\def\f{\phi}

\def\m{\mu}

\def\o{\omega}

\def\p{\pi}
\def\r{\rho}

\def\s{\sigma}

\def\t{\theta}

\def\F{\Phi}
\def\pa{\partial}

\def\to{\rightarrow}

\def\half{{1 \over 2}}
\def\sF{{{ F}\!\!\!\!\hskip.8pt\hbox{\raise1pt\hbox{/}}\,}}
\def\som{{{ \omega}\!\!\!\!\hskip.8pt\hbox{\raise1pt\hbox{/}}\,}}
\def\sJ{{{\rm J}\!\!\!\!\hskip.8pt\hbox{\raise1pt\hbox{/}}\,}}

\title{Chronology protection in stationary 3D spacetimes}
\author{Joris Raeymaekers
\\
Institute of Physics of the ASCR\\
Na Slovance 2, 182 21 Prague 8, Czech Republic}
\abstract{We study chronology protection in stationary, rotationally symmetric spacetimes in 2+1 dimensional gravity,
focusing especially on  the case of negative cosmological constant. We show that in such spacetimes  closed timelike curves
must  either exist all the way to the boundary or, alternatively, the matter stress tensor must violate the null energy condition in the bulk.
We also show that the matter in the closed timelike curve region gives a negative contribution to the conformal weight from the point of view of
the dual conformal field theory.  We illustrate these properties in a class of examples involving rotating dust in anti-de Sitter space,
and comment on the use of the AdS/CFT correspondence to study chronology protection.}
\preprint{arXiv:1106.5098 [hep-th]}
\keywords{AdS-CFT Correspondence, Classical Theories of Gravity}

\begin{document}
\section{Introduction}
The question whether  and how the laws of physics prevent the construction, in principle, of time machines
is a fascinating one
which goes to the heart of our understanding of spacetime geometry and quantum physics.
Hawking's chronology protection conjecture \cite{Hawking:1991nk} states  that the laws of physics prevent the formation
of closed timelike curves (CTCs) that would allow one to travel to one's past (see \cite{Thorne:1992gv,Visser:2002ua} for reviews and further references). At the classical level, evidence
for the conjecture comes from the fact that spacetimes with a compactly generated chronology horizon
require unphysical matter sources that violate the null energy condition. On the quantum level, the situation is less clear-cut.
In the semiclassical approximation, the stress-energy tensor can violate the null energy condition but this approximation is known to break down
in the presence of chronology horizons \cite{Kay:1996hj}.
Hence to settle the question definitively presumably requires a complete formulation where gravity itself is also quantized\footnote{In the context of string theory, dynamical mechanisms for resolving CTCs  have been proposed \cite{Drukker:2003sc,Raeymaekers:2010re}.}.

The case of asymptotically anti-de-Sitter (AdS) spacetimes provides a promising setting to address these thorny issues,  since the AdS/CFT
correspondence \cite{Maldacena:1997re} provides a definition of a quantum gravity theory on  AdS in terms of a conformal field theory (CFT) on the boundary.
The classical approximation in the bulk encodes a certain large $N$ limit of the boundary CFT.
Therefore gravity theories that have a CFT dual are constrained already on the classical level by physical requirements (such as unitarity) of the dual CFT.
 One would therefore expect to be able to identify unphysical classical solutions through some pathological behaviour of physical quantities, such as correlators, in the boundary CFT.
Indeed, several examples  have appeared in the literature \cite{Herdeiro:2000ap,Raeymaekers:2009ij} where CTCs in the bulk were linked to unitarity-violating one-point functions in the dual CFT. Furthermore,  AdS/CFT should allow one to address quantum corrections in the bulk systematically by taking into account $1/N$ corrections in the boundary theory.

Some of the simplest and most well-known examples of spacetimes with closed timelike curves don't have a compactly generated chronology horizon and therefore
fall outside of Hawking's original
argument. This is the case for `eternal' time-machines, where the space-time is stationary and every CTC is a member of an infinite  `tube' of CTCs formed by
time translating the original curve.
Examples include Van Stockum's 1937 solution \cite{vanStockum:1937zz} and the G\"odel universe \cite{Godel:1949ga}. In this paper, we will address the issue
of chronology protection in stationary spacetimes in the simplest context: we consider
the case of 2+1 dimensional gravity and for additional simplicity consider spacetimes which are also rotationally symmetric.
For the above mentioned reasons, we are especially interested in the AdS case although our arguments could be applied more generally.

A first question we will address is whether  such time machines require the  violation of energy conditions as in Hawking's argument. In stationary, rotationally
symmetric spacetimes, it suffices to consider
timelike azimuthal circles to which we will refer as azimuthal closed timelike curves (ACTCs). There are two classes of time machines depending on the behaviour of
the ACTCs, as illustrated in figure \ref{lcfig}.
\FIGURE{
\begin{picture}(400,90)
\put(-10,0){\includegraphics[width=200pt]{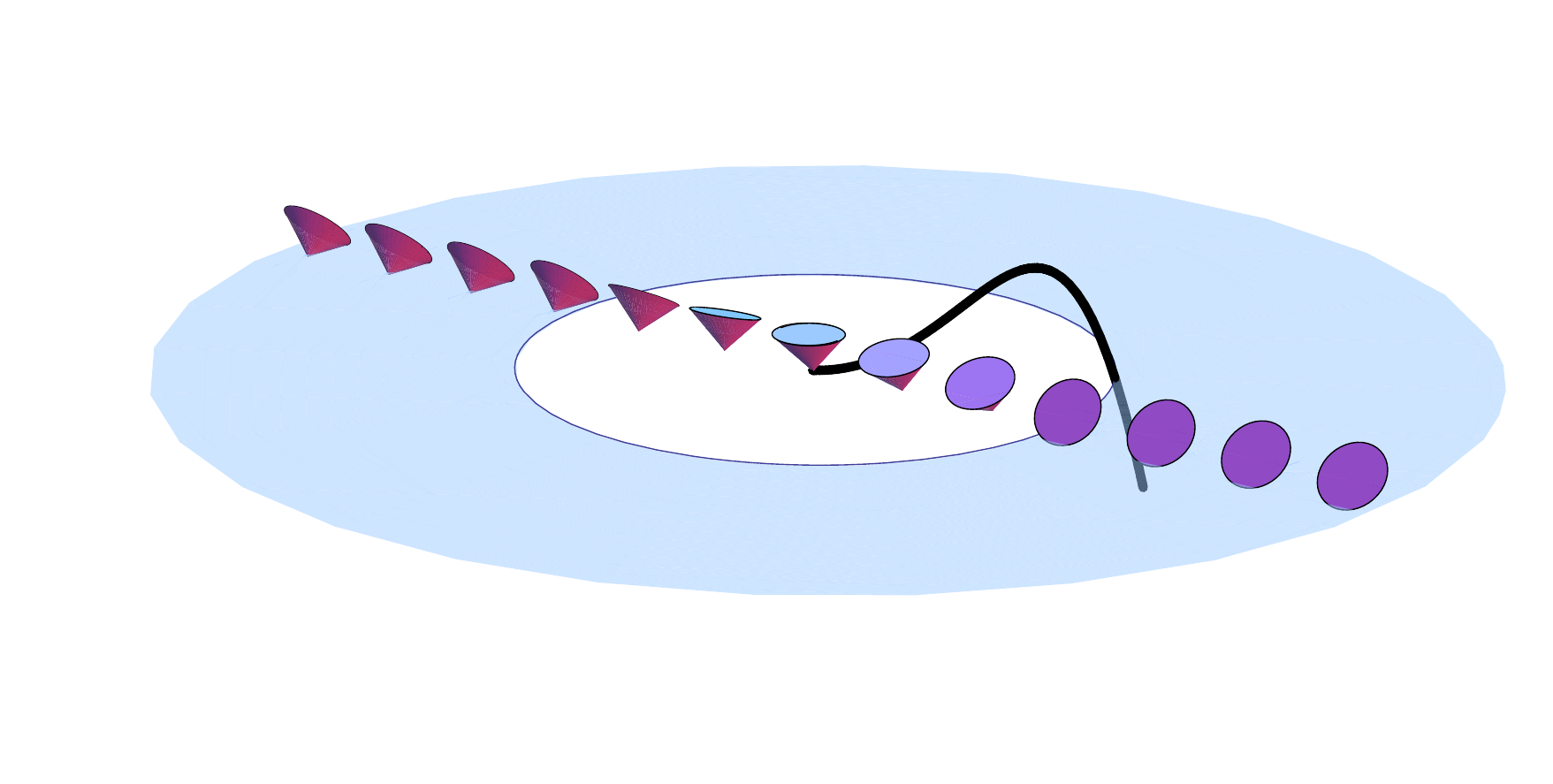}}
\put(180,-5){\includegraphics[width=240pt]{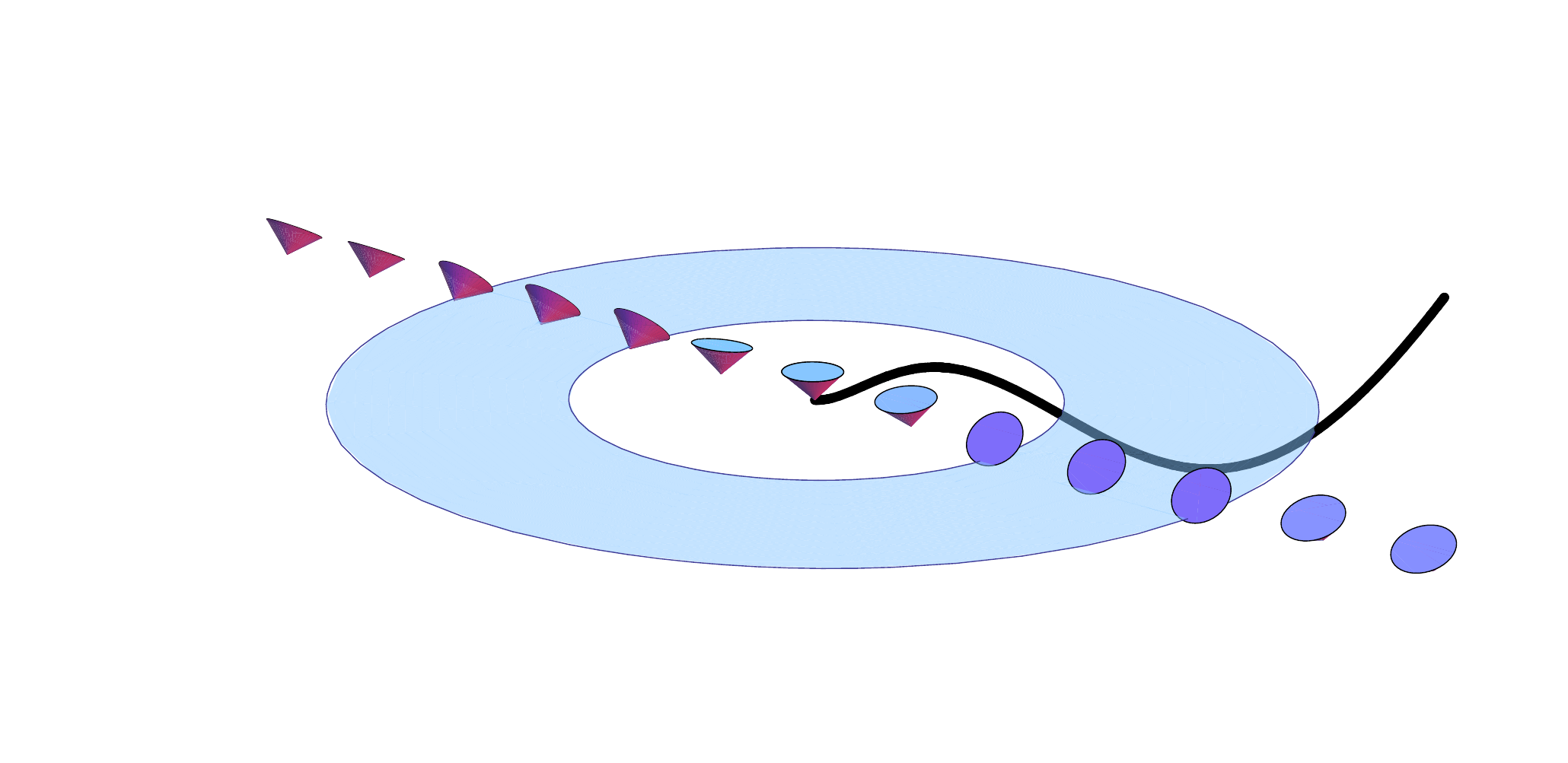}}
\end{picture}
\caption{The behaviour of local lightcones in  a constant time slice in two classes of time machines. The black curve is the length of the azimuthal Killing vector.
Left: the ACTCs persist all the way to the boundary. Right:
the ACTCS are localized in the constant time slice.}\label{lcfig}}

In the first class, the ACTCs persist all the way to the AdS boundary, in which case there no sensible way of associating a boundary CFT to the spacetime.
This in itself can of course be seen as an argument for the unphysical nature of this class of time machines.
An example within this class is the three-dimensional version of the G\"odel universe \cite{Rooman:1998xf}. A second class of spacetimes has  ACTCs
that don't reach the boundary and are localized in the constant time slice. In such spacetimes the light-cones tip over and subsequently `untip' in ACTC region of spacetime.
An example of this class of time machine was discussed in \cite{Raeymaekers:2009ij}.
We will that this tipping and untipping of the lightcones requires an unphysical matter source that violates the null energy condition in the ACTC region.
In the case of zero cosmological constant and sources which fall off fast enough at infinity, a similar result was obtained by Menotti and Seminara \cite{Menotti:1992nz}.
Our proof generalizes their result by extending  it  to the case of negative cosmological constant and, since it relies  only on local properties of the metric in the
ACTC region,  to
 sources which don't fall off at infinity.


A next issue we  address is in what precise sense the violation of the null energy condition  represents a negative contribution to the total energy of the spacetime.
In 3D asymptotically AdS spaces, there are two natural energies, namely the expectation values of the left- and right-moving conformal weights $L_0, \bar L_0$ in the
dual CFT.
We will find expressions for the total $L_0$ and $\bar L_0$ in time machines with localized ACTCs which make it clear the the ACTC region contributes
negatively to the total value of either $L_0$ or $\bar L_0$.

Lastly, we will  illustrate our general results in some examples involving rotating dust in AdS. We begin by revisiting the example of \cite{Raeymaekers:2009ij} involving a
 ball of G\"odel space embedded within AdS,
and clarify in particular how the NEC is violated when there are localized ACTCs. Since in this example, the total $L_0$ was found to be negative whenever there are CTCs,
one could wonder whether having localized ACTCs always implies a negative conformal weight. We show that this is not the case, by constructing a counterexample
which is completely smooth, has localized ACTCs and
has positive conformal weights.

\section{Chronology protection for stationary rotationally invariant metrics}
\subsection{Stationary rotationally symmetric metrics}
We consider 2+1-dimensional gravity
in the presence of a cosmological constant. Einstein's  equations are
 \be R_{ij} = 8\p G( T_{ij} - T g_{ij} ) + 2 \L g_{ij}.\label{Einsteq} \ee
 We  restrict attention to stationary, rotationally symmetric metrics, some consequences of
which we summarize here.
There
are  two commuting Killing vectors, $T$ and $\F$, where $T$ is
timelike and has open orbits and $\F$ has closed orbits.
We will consider both the axisymmetric case where $\F$  vanishes on a one-dimensional submanifold (the axis),
as well as cylindrical type universes without symmetry axis.
Away from the axis (if there is one), the spacetime is foliated by two-dimensional timelike
cylinders whose tangent space is spanned by $T,\F$. Taking $R$ to be the
unit normal to these cylinders, $T,\F$ and $R$
commute (see e.g. \cite{Deser:1986xf}) so that they define a coordinate basis $T = \pa_t, \F =
\pa_\f, R = \pa_\r$ in which the metric takes the form \be {ds^2 }
= d\r^2  + g_{tt} dt^2 + 2 g_{t\f} dt d\f + g_{\f\f} d\f^2
.\label{metransatz} \ee where $g_{tt}, g_{t\f}, g_{\f\f}$ are
functions of $\r$ only and $\f$ is identified modulo $2\p$.
In the case of axial symmetry we can assume that the axis
corresponds to $\r=0$ by
shifting the $\r$ coordinate.
The above considerations imply that
\bea
g_{tt} &<& 0\\
- \det g  &=& g_{t\f}^2 - g_{tt}g_{\f\f}   > 0  \label{posvol}
\eea
As for the analytical properties of the metric components, we will assume that they are continuous and differentiable,
but the derivative doesn't need to be continuous (in particular, we will allow for thin shells of matter in the spacetime).

We will study metrics of the above form which have closed timelike curves (CTCs).  The existence of closed timelike curves implies that $g_{\f\f}$ must become
negative for some value of $\r$ \cite{Menotti:1992nz}. Indeed,  any CTC will have a point where the value of the $t$ coordinate has a local extremum, such that the
tangent vector points lies in  the $\r,\f$ plane. Since the tangent vector is timelike, $g_{\f\f}$ at this point must be negative.
Hence the existence of CTCs implies the existence of timelike $\f$-circles, which we will call azimuthal closed timelike curves (ACTCs).


Our arguments will apply only to closed timelike curves located in regions of spacetime where the metric can be brought
in the form  (\ref{metransatz}), which in the presence of horizons is not possible globally.
For example, for BTZ black holes \cite{Banados:1992wn} in the $\L<0$ anti de-Sitter (AdS)  case, the metric can be brought in the form (\ref{metransatz})
outside of the black hole horizon, while in the  $\L >0$ de Sitter case the form (\ref{metransatz}) is valid only inside the cosmological horizon, and in this case
the coordinate system   (\ref{metransatz}) breaks down at finite $\r$.

For sufficiently localized
matter distributions, the metric will approach a vacuum solution
asymptotically. In these coordinates and for $\L \leq 0$, this means that the metric
approaches the AdS or Minkowski metric for $\r \to \infty$.

\subsection{A boost-free null triad}

Our argument will be simplified by expressing Einstein's equations (\ref{Einsteq}) in a conveniently chosen null triad $(e^+, e^-, e^\r )$ in terms of which the metric takes the form\footnote{Our conventions
are as follows: our signature is $-++$,
indices $i,j, \ldots$ refer to a coordinates basis and indices $a,b,\ldots$ refer to an orthonormal basis. In writing specific components, $t,\f$ refer to the
coordinate basis and $+,-$ refer to the triad basis.}
\be
ds^2 = - 2 e^+ e^- + (e^\r)^2 \equiv \h_{ab} e^a e^b.
\ee
In what follows we will only need  the $++$ and $--$  components of Einstein's equations (\ref{Einsteq}):
\be
R_{\pm\pm} = 8 \p G T_{\pm \pm} \label{einsteqlc}.
\ee
The components of the Ricci tensor can be written as
\bea
R_{\pm\pm} &=& e_\pm^i R_{ij} e_\pm^j\\
&=&  e_\pm^i (\nabla_j \nabla_i - \nabla_i \nabla_j) e_\pm^j\\
&=& \nabla_i v^i_\pm + (\nabla_i e^i_\pm)^2 - \nabla_j e^i_\pm \nabla_i e^j_\pm\label{Riccicomps}
\eea
where
\be
v^i_\pm \equiv e_\pm^j \nabla_j e_\pm^i - e_\pm^i \nabla_j e_\pm^j. \label{vdef}
\ee

We will now show that we can  choose our triad such that the last two terms in (\ref{Riccicomps}) are zero.
They can be expressed in terms of the Ricci rotation coefficients (i.e. the components of the spin connection one forms in the frame basis) defined as
\be
\o_{abc} = e_{cj} e^i_a \nabla_i e^j_b = - \o_{acb}.\label{omdef}
\ee
One finds
\be
(\nabla_i e^i_\pm)^2 - \nabla_j e^i_\pm \nabla_i e^j_\pm = 2 ( \o_{\r \mp \pm} \o_{\pm \r \pm } - \o_{\pm \mp \pm} \o_{\r\r \pm} ).
\ee

We will choose  $e^\r = d\r$ while $e^+,e^-$ have no component along $d\r$. This specifies the triad up to a single function
$B(\r )$:
\bea
e^\pm &=& {e^{\mp B} \over \sqrt{2}} \left[ \sqrt{-g_{tt} }dt - { g_{t\f} \pm \sqrt{-g} \over \sqrt{-g_{tt}}} d\f\right] \\
e^\r &=& d\r.\label{triad}
\eea
The function $B(\r ) $ parameterizes our freedom to  locally  Lorentz boost the triad; our arguments will be simplified by making a convenient choice for this function.
In this triad, one finds that some of the rotation coefficients are zero:
\be
\o_{\pm +-}=\o_{\r \r \pm }=0. \label{zerocoeffs}
\ee
Now consider the effect of a local boost transformation $B \to B + \l(\rho) $. The coefficient  $\o_{\r + -}$
transforms as a gauge potential\footnote{
The other coefficients transform as follows: $\o_{\pm\pm\r}$ scale with weight $\pm 2$ and
$\o_{\pm\mp\r}$ scale with weight 0.}
\be
\o_{\r + - } \to \o_{\r + - }  - \l'.
\ee
Hence by choosing the boost factor  $B$ in (\ref{triad}) judiciously we can also gauge away $\o_{\r + - }$, which means that our
triad vectors $e_\pm$ do not undergo a boost under parallel transport in the $\r$ direction\footnote{In the Newman-Penrose \cite{Newman:1961qr} formalism adapted to
3D gravity \cite{Hall:1987bz}, the properties (\ref{zerocoeffs})
mean that $\t =\tilde \t = \k=\tilde \k=0$ while our choice of boost parameter also gauges away $\e$. Note that $e_\pm$ are in general not tangent to a
geodesic null congruence, since
in general the coefficients $\o_{\pm\pm \r}$ ($\s, \tilde \s$ in the notation of \cite{Hall:1987bz}) are nonzero.}
In terms of the metric components, our gauge choice means that $B$ has to satisfy
\be
B ' =  {g_{tt} g_{t\f}' - g_{t\f} g_{tt}'\over 2 g_{tt} \sqrt{-g} }\label{Beq}
\ee
In view of (\ref{posvol}) this differential equation is regular in the region of interest and hence we can always construct a triad with the desired properties.

In the triad (\ref{triad},\ref{Beq}), the $++$ and $--$ components of Einstein's equations (\ref{einsteqlc}) become simply
\be
\nabla_i v_\pm^i = 8\p G T_{\pm \pm}
\ee
which, since everything depends only on $\r$, can be written as
\be
\pa_\r \left( \sqrt{-g} v_\pm^\r \right) = 8\p G \sqrt{-g} T_{\pm \pm} \label{Komarid}.
\ee
The left hand side is a total derivative while the right-hand side is positive when the NEC holds. This identity will be our main tool in what follows.

Using (\ref{vdef},\ref{omdef},\ref{zerocoeffs})  $v_\pm^\r$ is easily seen to be equal to a Ricci rotation coefficient:
\be
v_\pm^\r = \o_{\pm\r\pm}
\ee
and, for later use, we give the explicit expression for $\sqrt{-g} v_\pm^\r$:
\be
\sqrt{-g} v_\pm^\r = {e^{\pm 2 B} \over 4 \sqrt{-g}}\left( g_{tt} g_{\f\f}' - g_{\f\f} g_{tt}' +2 {g_{t\f} \mp \sqrt{-g} \over g_{tt} }\left( g_{t\f} g_{t t}' - g_{t t} g_{t \f}'\right)\right).\label{vrho}
\ee

\subsection{Chronology protection from the null energy condition}\label{chronprotnec}
Now we will consider spacetimes which contain ACTCs. It will be useful to distinguish two possibilities for the behavior of the ACTCs.

The first possibility is that  the ACTCs are present all the way to $\r \to \infty$. Such metrics have pathological asymptotic behaviour,
and one should presumably  discard them by imposing suitable physical boundary conditions \cite{Hawking:1991nk}. For example, in the $\L <0$ case,
it seems unlikely that one could associate a sensible CFT to a  spacetime with CTCs on the boundary.

Hence we will focus on a the second possibility, where there are no CTCs for $\r \to \infty$. This means that $g_{\f\f}$  must have a zero at a radius $\r_+$ where it goes from being negative for slightly smaller radii to being positive for slightly larger ones.
We will further assume  that $g_{\f\f}$ has a second zero at a radius $\r_- < \r_+$ where it goes from being positive to being negative. This assumption can be be made without much loss of generality for the following reason.
For spacetimes
with a regular axis of symmetry, there must always be such a $\r_-$ since $g_{\f\f}$ is positive in the vicinity of the axis, as one can see by choosing local inertial coordinates on the axis (see \cite{Mars:1992cm} for a rigorous proof). If there is no axis of symmetry and no radius  $\r_-$, the metric has wormhole-like behaviour with a second asymptotic region for $\r \to - \infty$ where $g_{\f\f}$ is negative, and hence it belongs to the first class of metrics.

We will show that these time machines require an unphysical stress-energy tensor that violates the null energy condition (NEC) in the ACTC region between $\r_-$ and $\r_+$.
We recall that the NEC requires that \be k^i T_{ij} k^j \geq 0\label{NEC}\ee for any null vector $k$.

First let's summarize the above discussion of the behavior of $g_{\f\f}$ in the interval $[ \r_-, \r_+ ]$ (see also Figure \ref{gffvrhofig}):
\bea
g_{\f\f}(\r_-) &=& g_{\f\f}(\r_+) =0\\
g_{\f\f}(\r) &\leq & 0 \qquad for \ \r_-\leq\r \leq \r_+\\
g_{\f\f}'(\r_-) &<&0\\
g_{\f\f}'(\r_+) &>&0
\label{rhoplmin}
\eea
\FIGURE{
\begin{picture}(300,125)
\put(0,0){\includegraphics[width=200pt]{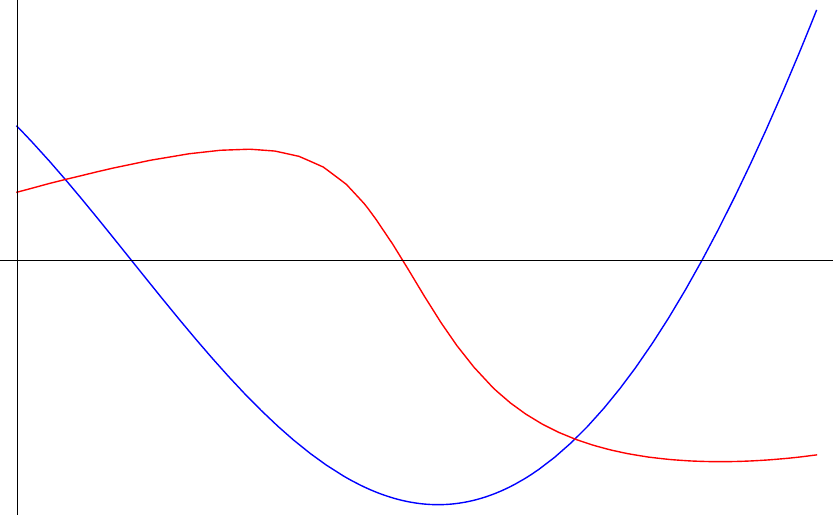}}
\put(20,53){$\r_-$}
\put(85,53){$\r_c$}
\put(151,53){$\r_+$}
\end{picture}
\caption{An example of the behavior of $g_{\f\f}$ (in blue) and $\sqrt{-g} v^\r_\s$ (in red) in a spacetime with localized ACTCs}\label{gffvrhofig}}
This behavior implies some properties of $g_{t\f}$ which we will need later. First of all,  $g_{t\f}$ cannot be zero in $\r_-$ or $\r_+$ because
of (\ref{posvol}).
Furthermore,   $g_{t\f}$ cannot change sign in the interval  $[ \r_- , \r_+ ]$. Indeed, if it did,
it would have a zero somewhere in the interval, and at that radius the metric would not be Minkowskian.
So we can consistently define
\be
\s \equiv {\rm sign}\, g_{t\f} (\r_- ) = {\rm sign}\, g_{t\f} (\r_+ )\label{sigmadef}
\ee
Now, when we evaluate the quantity $\sqrt{-g} v_\s^\r$ in  $\r_\pm$ using (\ref{vrho}), we see that  only the first term is nonzero:
\be
\sqrt{-g} v_\s^\r (\r_\pm ) = {e^{2 \s B} \over 4 \sqrt{-g}} g_{tt} g_{\f\f}' (\r_\pm).
\ee
Now we can integrate (\ref{Komarid}) between the radii $\r_-$ and $\r_+$ in (\ref{rhoplmin}) to get our main identity
\be
8\p G \int_{\r_-}^{\r_+} d\r \sqrt{-g}  T_{\s\s}  = {e^{2 \s B} \over 4 \sqrt{-g}} g_{tt} g_{\f\f}' (\r_+ )- {e^{2 \s B} \over 4 \sqrt{-g}} g_{tt} g_{\f\f}' (\r_- ) <0.\label{mainid}
\ee
The last inequality follows from (\ref{rhoplmin}).
Hence we see that the  null energy condition must be violated on average in the CTC region by an amount specified by the right hand side. We can also write a slightly more refined
estimate of NEC violation by observing that, since $\sqrt{-g} v_\s^\r$ changes sign between $\r_-$ and $\r_+$, there must be a radius $\r_c$ where (see Figure \ref{gffvrhofig})
\be
\sqrt{-g} v_\s^\r (\r_c)=0. \label{rhocdef}
\ee
We then see that the averaged null energy condition must be violated in the intervals $[ \r_-, \r_c ]$ and $[ \r_c, \r_+ ]$
separately:
\bea
 8\p G \int_{\r_-}^{\r_c} d\r \sqrt{-g}  T_{\s\s}  &=& - {e^{2 \s B} \over 4 \sqrt{-g}} g_{tt} g_{\f\f}' (\r_- ) <0\\
8\p G  \int_{\r_c}^{\r_+} d\r \sqrt{-g}  T_{\s\s}  &=& {e^{2 \s B} \over 4 \sqrt{-g}} g_{tt} g_{\f\f}' (\r_+ )<0 .\label{mainid2}
\eea

Having established this result, we can compare with the existing result in the literature \cite{Menotti:1992nz} for vanishing cosmological constant.
In this case one can show, by integrating
(\ref{Komarid}) between any radius $\r_0$ and $\r = \infty$ and sufficiently fast falloff of the sources at infinity, that  $g_{\f\f} /\sqrt{-g}$ must be a
nondecreasing function of $\r$ if the NEC holds and hence no CTCs can develop. One can easily extend this argument to  negative
cosmological constant and show that $g_{\f\f} /\sqrt{-g}$ is a nondecreasing function if the NEC holds and the spacetime is asymptotically
conical (meaning that both $L_0 \leq c/24$ and   $\bar L_0 \leq c/24$).
However, for spacetimes which are not asymptotically conical, $g_{\f\f} /\sqrt{-g}$ can be decreasing even if the NEC is obeyed, as we will
see in the example in section \ref{nonsing}.
Our argument leading to (\ref{mainid}) holds  for these spacetimes as well and, since it only relies on the properties of the metric in the ACTC region, doesn't
require any restrictions on the falloff of the sources.
\subsection{CTCs and conformal weights}
One expects intuitively that the violation of the NEC in the CTC region represents a negative contribution to the total energy of the system.
In this section we will make this more precise in the case of negative cosmological constant:
we will see that the CTC region represents a negative contribution to either $L_0$ or $\bar L_0$.

First we evaluate the asymptotic values of the quantities $\sqrt{-g} v_\pm^\r$ in the cases where the matter stress tensor vanishes sufficiently fast
for $\r \to \infty$. The metric then asymptotically approaches a vacuum solution
which is stationary and rotationally invariant; the general solution of this kind involves two integration constants which can be identified with the mass and
 angular momentum
\cite{Banados:1992wn}. In Schwarzschild-like coordinates the metric takes  the form
\be
{ds^2 } = -  \left(- M - \L r^2 \right) d   t ^2  - J  d  t d  \f + r^2 d  \f^2 + {d r^2 \over - M - \L r^2 + {J^2 \over 4 r^2}}\label{BTZ}
\ee
The mass and angular momentum are related to the conformal weights in the dual CFT as
\bea
L_0  &=&{c \over 24}( M+ J/L+1)\\
\bar L_0 &=& {c \over 24}( M-  J/L +1)\label{confweights}
\eea
where $c$ is the Brown-Henneaux central charge \cite{Brown:1986nw} and where we have set
\be
\L = -{1\over L^2}.
\ee
The transformation to the proper radial coordinate $\r$ is
\be
r^2 = {L^2 \over 2 } \left( d^2  e^{2 \r/L} +  { M}  \right) + \calo (e^{- 2 \r/L} )\label{schwarztoproper}
\ee
with $d^2$ an integration constant, and the metric near $\r \to \infty$ takes the Fefferman-Graham \cite{FG} form
\be
{ds^2 } = d\r^2 +  { d^2 e^{2 \r/L}\over 2 }  \left(-  d   t ^2  + L^2 d\f^2 \right) + { M \over 2} dt^2 - J dt d\f + { M \over 2} L^2 d\f^2 + \calo (e^{- 2 \r/L} )\label{asads}
\ee
Note that to bring a general metric into this form, in which the boundary metric is diagonal, we need to make a coordinate transformation of the form
$t\to c_1 t, \ \f \to \f + c_2 t $ which we have left unfixed so far.
The equation (\ref{Beq})  near the boundary implies that $B$ goes to a constant
\be
B = B_\infty + \calo (e^{- 2 \r/L} )
\ee
and for the boundary behavior of $\sqrt{-g} v_\pm^\r$ we find
\be
\sqrt{-g} v_\pm^\r = {e^{\pm 2 B_\infty} \over 2} \left( M \pm {J \over L} \right) + \calo (e^{- 2 \r/L} )\label{vrhobound}
\ee
Of course, for the exact vacuum solution  (\ref{BTZ}), we know from
(\ref{Komarid}) that $\sqrt{-g} v_\pm^\r$ should be independent of $\r$, and one easily checks that
the leading approximation (\ref{vrhobound}) becomes exact in this case.
We conclude that, if we impose the following boundary condition on (\ref{Beq})
\be
\lim_{\r \to \infty} B (\r) = 0 \label{bcB}
\ee
the asymptotic  values of $\sqrt{-g} v_+^\r $ and $\sqrt{-g} v_-^\r $ measure essentially the left- and right-moving conformal weights.

We now use this to give a convenient expression for the conformal weight for spaces with localized CTCs.
Integrating  (\ref{Komarid}) between $\r_c$ and infinity and using (\ref{vrhobound}) we get, e.g. for $\s = +1$:
\be
{ 24\over c }  L_0 = 2 \int_{\r_c}^{\r_+} d\r \sqrt{-g}  T_{++} + 2 \int_{\r_+}^{\infty} d\r \sqrt{-g}  T_{++} + 1   \label{enads}
\ee
The first term on the RHS comes from the CTC region and is negative as we argued already in (\ref{mainid2}). In this precise sense, the CTC
region represents a negative contribution to the total left- or right-moving conformal weight.

\section{Examples: rotating dust solutions in AdS}

We will now illustrate the  above properties with examples. We will find a simple class of analytic solutions involving rotating dust in AdS.
These generalize the three-dimensional G\"odel solution and can be engineered to have localized CTC regions. We take units where
\be
\L = - { 1 \over 4}
\ee
and consider metrics of the form
\be
{ds^2 } = d\r^2 -d\tilde t^2 + 2 l d\tilde t d\tilde \f + (l'^2 - l^2) d\tilde \f^2
\ee
for some function $l(\r)$. Since the determinant of the metric is $|l'|$,  $l'$ cannot vanish (except on the symmetry axis if there is one), and hence
it must have the same sign everywhere.
If the spacetime has a regular axis of symmetry, which we can take to be at $\r = 0$, we can
 impose that $(\tilde t, \tilde \f, \r)$  are  local inertial
coordinates along the axis, leading to the conditions
\bea
l(0) &=& 0\\
l'(0) &=& 0\\
l''(0) &=& 1.\label{smoothaxis}
\eea

These metrics satisfy the Einstein equations with a pressureless, rotating dust source:
\be
T^{ab} = R u^a u^b.
\ee
where the energy density $R$ is given in terms of $l$ as
\be
R = 1 - {l''' \over l'}.\label{Req}
\ee
The velocity vector is $u = \pa_{\tilde t}$, so this coordinate system is comoving with the dust.
Within this class of metrics we can construct analytic solutions with the desired behavior of $g_{\f\f}$ by finding a suitable function $l$.

For asymptotically AdS solutions, the solution should approach a vacuum metric with $R=0$ for large $\r$.
The function $l$ then behaves  for large $\r$ as
\be
l = \e \left( b_1^2 e^\r - {a^2} + {b_2 \over 4} e^{-\r} + \calo (e^{- 2 \r} ) \right)\label{asl}
\ee
where $\e$ is a sign factor, $\e = {\rm sgn} (l' )$.
The above comoving coordinates ($\tilde t, \tilde \f, \r$) are in general not the ones in which the metric takes the asymptotic form (\ref{asads}).
The transformation to the asymptotic AdS coordinates $t,\f$ in  (\ref{asads}) is
\bea
\tilde t &=& a^2 t\\
\tilde \f &=& \f - \e {t\over 2}
\eea
and comparing to (\ref{asads}) we can read off the mass and angular momentum:
\bea
M &=& - {a^4 + b_1^2 b_2 \over 2}\\
{J\over L} &=& \e {a^4 - b_1^2 b_2 \over 2}\label{MJdust}
\eea
Using (\ref{confweights}), we  find the  conformal weights in the dual CFT for $\e = 1$:
\bea
{24 \over c} L_0 &=& M + {J\over L} + 1 = 1 - b_1^2 b_2\\
{24 \over c} \bar L_0 &=& M - {J\over L} + 1 = 1 - a^2.\label{L0ab}
\eea
For $\e=-1$, the roles of $L_0$ and $\bar L_0$ are reversed.

The solutions for $B$ with boundary condition (\ref{bcB}) and  for$\sqrt{-g} v_\pm^\r$ are also easily found in the $(t,\f,\r)$ coordinates.
For $\e = 1$ one finds
\bea
\sqrt{-g} v_+^\r &=&  b_1^2 e^{ \r}( l'-  l'')\\
 \sqrt{-g} v_-^\r &=& -{a^4 \over 4 b_1^2}  e^{ -\r}( l' +  l'')\label{vs}
\eea
and one checks that $\sqrt{-g} T_{\pm\pm}$ are indeed obtained by taking a radial derivative in accordance with (\ref{Komarid}).
Furthermore one can also check that the asymptotics go as (\ref{vrhobound}) by plugging in the  expansion (\ref{asl}).
When $\e = -1$, one has to replace $\sqrt{-g} v_\pm^\r \to - \sqrt{-g} v_\mp^\r$ in these expressions.

\subsection{The G\"odel dust ball}\label{dustball}
The simplest example of this kind is obtained by taking the energy density of the dust to be constant.
It's convenient to parameterize
\be
R = {\m -1\over \m}\label{Rconst}
\ee
with $\m $ a constant greater than one. The solution of (\ref{Req}) with boundary conditions (\ref{smoothaxis}) is
\be
l =  \m \left( 1 - \cosh  { \r \over  \sqrt{\m}}\right)\label{Godell}
\ee
This solution is known as the 3D G\"odel universe \cite{Rooman:1998xf}, because for the particular value $\m = 2$ it describes the nontrivial
three-dimensional part of G\"odel's original solution \cite{Godel:1949ga}. For $\m, =1$ one recovers the  AdS metric in global coordinates.
This metric has two special radii.
 At the radius
\be \r_B =\sqrt{\mu } {\rm arccosh}\left(\frac{\mu }{\mu -1}\right) \label{rBdef}\ee
the metric coefficient $g_{\f\f} $ is maximal; beyond this radius it decreases with $\r$\footnote{The radius $\r_B$ is the locus where the expansion of a congruence of null geodesics,
emitted from the origin, becomes zero. It plays a special role as a preferred holographic screen in Bousso's covariant holography proposal \cite{Bousso:1999cb,Boyda:2002ba}.}.
At the radius
\be \r_{CTC} = \sqrt{\m} {\rm arccosh } \left({  \m  + 1\over  \m -1}\right) \label{rctcdef} \ee
the coefficient $g_{\f\f}$ is zero; beyond this radius it becomes negative and there are ACTCs.
This is an example of the first type of time machine discussed in section \ref{chronprotnec}: the null energy condition is satisfied everywhere but the CTCs persist to infinity.

One way of obtaining an asymptotically AdS solution is to take a finite ball of dust with constant density $R$ up to some radius
$\r_0$. The solution for $\r \geq \r_0$ is then a vacuum solution determined by solving the Israel matching conditions \cite{Israel:1966rt} at the edge of the dust ball $\r = \r_0$.
This is the example considered in \cite{Lubo:1998ue,Raeymaekers:2009ij}  which we will here expand on and clarify.
 We will consider two types of matching: one without and one
with a singular shell of matter  at the matching surface $\r = \r_0$.
We will see that, without including a  singular shell, matching onto an asymptotically AdS metric is only possible for radii
$\r_0 \leq \r_B$ defined in (\ref{rBdef}). 
For $\r_0 > \r_B$, matching onto asymptotically AdS is only possible when including a singular shell of matter at the matching surface; and we will see
that  it is this singular source which violates the NEC when we take $\r_0 \geq \r_{CTC}$ and the configuration has CTCs.

\subsubsection*{Matching without singular shell}
First we consider the matching problem without singular shell. For $\r \geq \r_0$, the function $l$  must be a solution of (\ref{Req}) with $R =0$,
and matching conditions reduce to requiring continuity of $l$ and  it's first and second derivatives at $\r = \r_0$.
This leads to the dust ball  solution
\bea
l &=&  \m \left( 1 - \cosh  { \r \over  \sqrt{\m}}\right) \qquad {\rm for }\ \r\leq \r_0 \\
l &=& \cosh {\r_0\over \sqrt{\m}} \left( \m - 1 + \cosh (\r - \r_0) \right) + \sqrt{\m} \sinh  {\r_0\over \sqrt{\m}}\sinh (\r - \r_0)-\m \ \  {\rm for }\ \r\geq \r_0 .
\eea
In the exterior region $l$ is of the form (\ref{asl}) with
\bea
\e &=& 1\\
b_1^2 &=& \frac{1}{2} e^{-\r_0} \left(\sqrt{\mu } \sinh \left(\frac{\r_0}{\sqrt{\mu }}\right)+\cosh
   \left(\frac{\r_0}{\sqrt{\mu }}\right)\right)\\
a^2 &=& \m - (\m - 1) \cosh {\r_0\over \sqrt{\m}}\\
b_2 &=& 2 e^{\r_0} \left(\cosh \left(\frac{\r_0}{\sqrt{\mu }}\right)-\sqrt{\mu } \sinh
   \left(\frac{\r_0}{\sqrt{\mu }}\right)\right)
\label{abnonsing}
\eea
and from (\ref{L0ab}) we  read off the conformal weights
\bea
{24 \over c} L_0 &=& (\m - 1) \sinh^2 {\r_0\over \sqrt{\m}} \\
{24 \over c} \bar L_0 &=& 1 - \left(  \m - (\m - 1) \cosh {\r_0\over \sqrt{\m}} \right)^2.
\label{matchingsol1}
\eea
Some properties of this matched solution are illustrated in figure (\ref{nonsing}).
\FIGURE{
\begin{picture}(370,100)
\put(-10,0){\includegraphics[width=120pt]{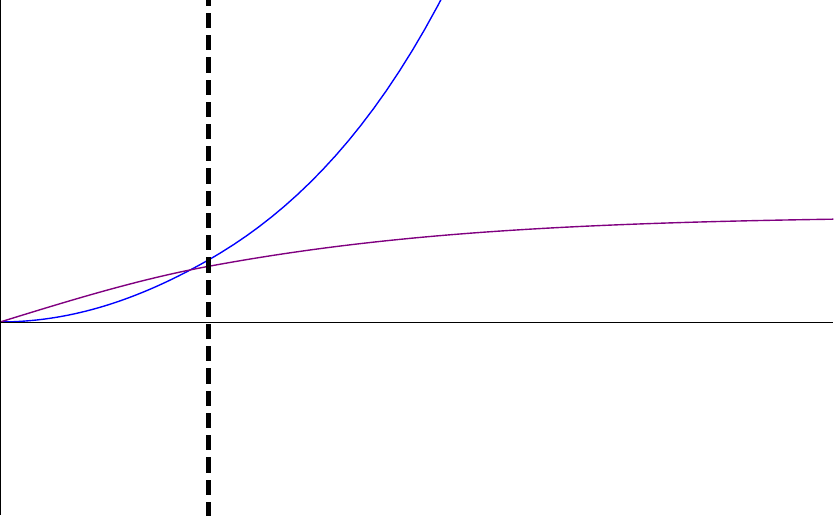}}
\put(120,0){\includegraphics[width=120pt]{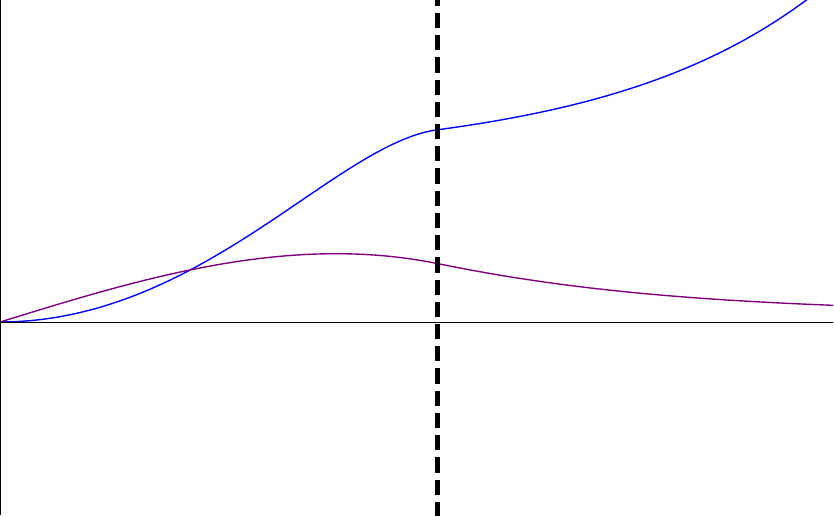}}
\put(250,0){\includegraphics[width=120pt]{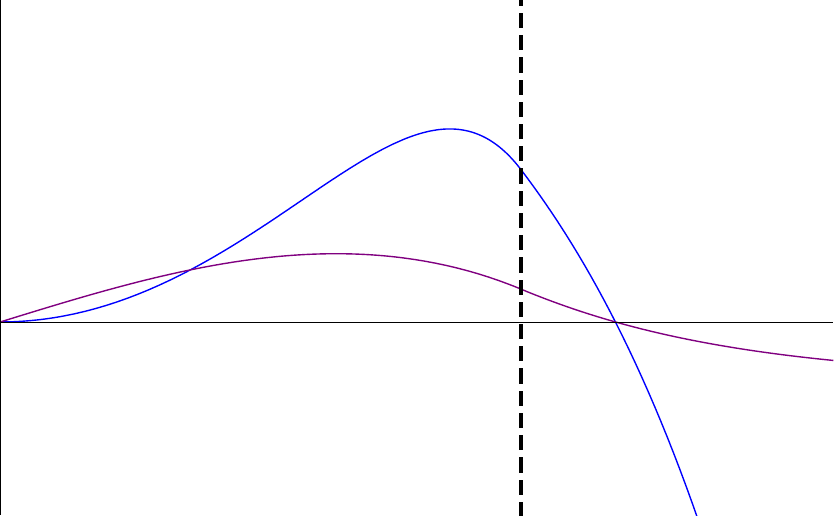}}
\put(40,-10){$\r_0 < \r_t$}
\put(150,-10){$\r_t < \r_0 < \r_B$}
\put(290,-10){$ \r_0 > \r_B$}
\end{picture}
\caption{The G\"odel dust ball discussed in the text for various values of $\r_0$. The blue curve is $g_{\f\f}$, the purple one is $g_{\f\f}/\sqrt{-g}$ and the dotted line denotes the edge $\r_0$ of the dust ball.}\label{nonsing}}
For small $\r_0$, the outside metric is that of a spinning conical defect with $M^2 + \L J^2 >0 $ and $-1 \leq M <0$.
At a certain value of the matching radius,   \be \r_t= \sqrt{\m} {\rm arccosh} \sqrt{\m \over \m -1}, \ee
 the metric  crosses over into the overspinning
regime with $M^2 + \L J^2 <0$.  Increasing $\r_0$ further, something special happens when $\r_0 >\r_B$: from (\ref{abnonsing}) we see that $a^2$ becomes negative. The metric behaves asymptotically
as in (\ref{asads}) with $d^2 = a^2 b_1^2$, which implies
that $g_{\f\f}$ becomes negative for large $\r$. Since there are CTCs at $\r \to \infty$, the metric is not asymptotically AdS. How could it happen that we have found a vacuum solution
on the outside of the dustball which is not asymptotically AdS? The answer becomes clear when we look  at the transformation to  Schwarzschild-like coordinates
 given by  (\ref{schwarztoproper}).
 When $a^2$ is negative, $\r \to \infty$ corresponds
to $r^2 \to - \infty$, which means that the outside metric  is a negative $r^2$ continuation of a spinning conical defect metric. This probably means that the dust ball
becomes unstable against gravitational collapse at $\r_0 = \r_B$. This is also suggested by the fact that as $\r_0$ approaches $\r_B$, we reach  the extreme black hole limit with $M = - J/L$ and  positive $M$\footnote{To be
precise, as $\r_0 \to \r_B$, the outside metric becomes a near-horizon scaling limit of the extremal BTZ black hole metric known as the null orbifold
\cite{Coussaert:1994tu}, as explained in \cite{Raeymaekers:2010re}.}.

\subsubsection*{Matching with a singular shell}
Above we saw that, without singular sources on matching surface, the G\"odel ball with $\r_0 > \r_B$ does not match to an asymptotically
AdS space. We can however match to asymptotically AdS by including a suitable singular source.
There is a more or less canonical source one can include: from the above remarks we see that we could match to to an asymptotic AdS space if, on the outside, we could replace
$\r\to -\r +  c$ without spoiling continuity of the metric. This is the case for $c = 2 \r_0$, so we are led to the configuration
\bea
l &=&  \m \left( 1 - \cosh  { \r \over  \sqrt{\m}}\right) \qquad {\rm for }\ \r\leq \r_0 \\
l &=& \cosh {\r_0\over \sqrt{\m}} \left( \m - 1 + \cosh (\r - \r_0) \right) - \sqrt{\m} \sinh
{\r_0\over \sqrt{\m}}\sinh (\r - \r_0)-\m \ \  {\rm for }\ \r\geq \r_0 .\label{dustballsing}
\eea

For this configuration, $l'$ and hence also the extrinsic curvature\footnote{In our coordinate system one has $K_{ij} = \half g'_{ij}$.} $K_{ij}$ changes sign across the matching surface, which
implies that  there is a singular source there. Since the
generic matching condition reads
\be
K^+_{ij} = K^-_{ij} - (T^s_{ij} - T^s g_{ij}),
\ee
our configuration corresponds to the singular source
\be
T_{ij}^s = 2 (K_{ij}^- - K^- g_{ij} ).\label{singsource}
\ee

The outside configuration (\ref{dustballsing}) is of the form (\ref{asl}) with
\bea
\e &=& - 1\\
b_1^2 &=& \frac{1}{2} e^{-\r_0} \left(\sqrt{\mu } \sinh \left(\frac{\r_0}{\sqrt{\mu }}\right)-\cosh
   \left(\frac{\r_0}{\sqrt{\mu }}\right)\right)\\
a^2 &=& (\m - 1) \cosh {\r_0\over \sqrt{\m}}-\m \\
b_2 &=& - 2 e^{\r_0} \left(\cosh \left(\frac{\r_0}{\sqrt{\mu }}\right)+\sqrt{\mu } \sinh
   \left(\frac{\r_0}{\sqrt{\mu }}\right)\right).
\label{absing}
\eea
For $\r_0 \geq \r_B$, one checks that the combination $a^2 b_1^2$ is now positive, and that the metric is asymptotically AdS.
For the values of the conformal weights one finds
the same expressions as in the case without singular source, but with $L_0$ and $\bar L_0$ interchanged:
\bea
{24 \over c}  L_0 &=& 1 - \left(  \m - (\m - 1) \cosh {\r_0\over \sqrt{\m}} \right)^2\\
{24 \over c} \bar L_0 &=& (\m - 1) \sinh^2 {\r_0\over \sqrt{\m}} .
\label{matchingsolsing}
\eea

The behavior of various quantities in this example are plotted in Figure \ref{singfig}.
\FIGURE{
\begin{picture}(370,100)
\put(-10,0){\includegraphics[width=120pt]{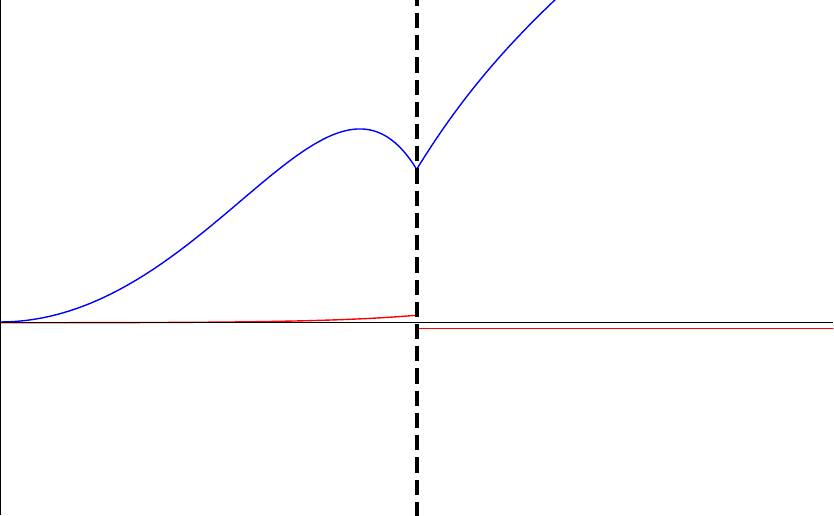}}
\put(130,0){\includegraphics[width=120pt]{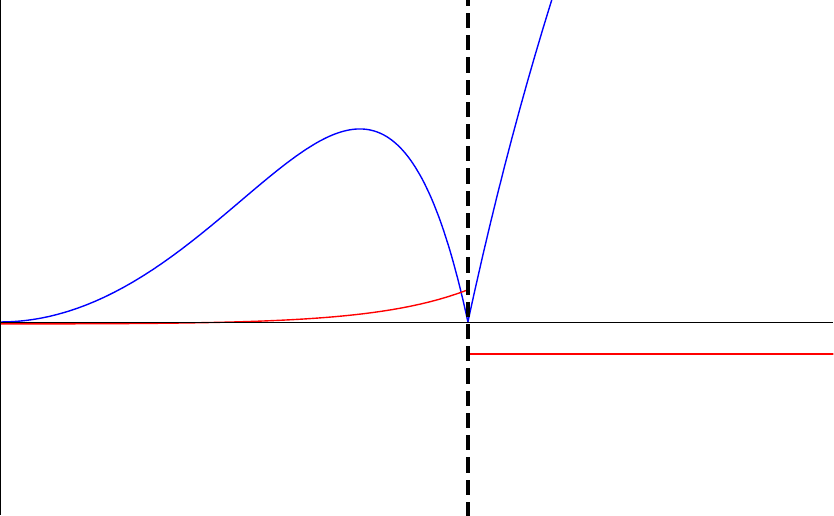}}
\put(270,0){\includegraphics[width=120pt]{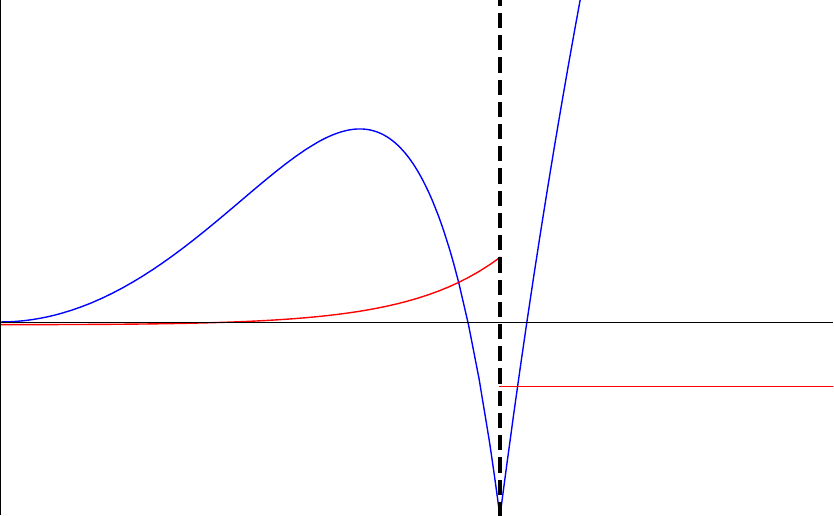}}
\put(40,-10){$\r_B <\r_0 < \r_{ctc}$}
\put(160,-10){$\r_0 =\r_{ctc}$}
\put(310,-10){$\r_0 >\r_{ctc}$}
\end{picture}
\caption{The G\"odel dust ball  with a singular source on the edge  for various values of $\r_0$. The blue curve is $g_{\f\f}$, the
red one is $\sqrt{-g} v_+^\r$ and the dotted line denotes the edge $\r_0$.}\label{singfig}}
For $\r_0 > \r_{ctc}$, the matched configuration has localized ACTCs\footnote{
In fact, for $\r_0 \geq \r_B$  these were the matched solutions  considered in \cite{Raeymaekers:2009ij} although  the presence of the singular source was overlooked there.}.
 Our argument of section \ref{chronprotnec} tells us that the null energy condition should be violated in the CTC region. It is well-known that the stress tensor of the interior G\"odel
space doesn't violate any energy conditions, so the NEC violation must come from the
 thin shell contribution at the edge. Let's make this more precise and show that our main identity (\ref{mainid}) is indeed satisfied in this example.

To verify our main identity (\ref{mainid}), we should take into account the delta function  term in the stress-energy, obtaining
\be
8\p G \int_{\r_{ctc}}^{\r_0} d\r \sqrt{-g}  T_{++}  + 8\p G \sqrt{-g}  T_{++}^s (\r_0) = {e^{2  B} \over 4 \sqrt{-g}} g_{tt} g_{\f\f}' (\r_+ )- {e^{2  B} \over 4 \sqrt{-g}} g_{tt} g_{\f\f}' (\r_- ) <0.\label{checkid}
\ee
Here we have taken the sign $\s = +1$ in (\ref{mainid}), since $l$ is positive in the ACTC region.
This equation can be checked using the derived formulas; in the exterior region we can use (\ref{vs}) (note that $\e = -1$ there), and for the
 interior quantities we need to impose that $B$  should be continuous across the matching surface; the function $\sqrt{-g} v^\r_+ $ changes sign there. For the left hand side one finds:
\bea
8\p G \int_{\r_{ctc}}^{\r_0} d\r \sqrt{-g}  T_{++} &=& {a^4 \over 2} \left( 1 + { 1\over h(\r_0)}\right)\\
8\p G \sqrt{-g}  T_{++}^s (\r_0) &=& - a^4\label{Ts}
\eea
where we defined
\be
h (\r) = \sqrt{\m} \sinh {\r \over \sqrt{\m}} - \cosh {\r \over \sqrt{\m}}.
\ee
Since $h$ is positive, the contribution coming from the G\"odel stress tensor is positive as expected. The contribution from the singular term
is negative, and such that the sum ${a^4 \over 2} \left(  { 1\over h(\r_0)} - 1\right)$ is also negative, since $h(\r_0)$ is greater
than one for $\r_0> \r_{ctc}$. We can also verify the conformal weight formula (\ref{enads}), where $\r_c$ has to be taken to be the matching radius
where  $\sqrt{-g} v^\r_+ $ changes sign. The formula  (\ref{enads}) then becomes
\be
{24 \over c} L_0 = T^s_{++} + 1
\ee
 which is indeed satisfied.

The current example has the intriguing property that the total $L_0$ becomes negative precisely when the spacetime contains CTCs; this is because
the parameter $a$ becomes one precisely  when $\r_0= \r_{ctc}$. From the point of AdS/CFT, such spacetimes are unphysical because unitarity of the dual CFT forbids
negative values of $L_0$.
Since we saw that our example involved a  tuned source on the matching surface, a natural question to ask is whether this behavior is generic.
In the following example we will see that the answer is in the negative: there exist spacetimes with localized ACTCs which have positive conformal weights.

\subsection{A smooth solution with localized CTCs}\label{smoothexsection}
In this example we display a simple class of smooth solutions which are asymptotically AdS and have localized CTCs.
We take the function $l$ to be of the form
\be
\ln l = \r - {a^2} e^{-\r} + {b - 2 a^4 \over 4} e^{- 2\r} + ( c_1 + c_2 \r + c_3 \r^2)  e^{- 3\r}
\ee
The first three terms guarantee the proper asymptotic behaviour (\ref{asl}), with the constants $a,b$ related to the total mass and angular momentum by (\ref{MJdust}).
The coefficients $c_i$ in the fourth term can then be chosen such that $g_{\f\f}$ has two zeroes for small $\r$.

As an illustrative example, we  take
\bea
a^2 &=& b = \half\\
c_1 &=& - {1 \over 2} \qquad c_2 =3 \qquad c_3 = -10
\eea
From (\ref{L0ab}) we see that this metric has positive conformal weights and asymptotes to a spinning conical defect:
\be
{24 \over c} L_0 = \half \qquad {24 \over c} \bar L_0 = {3\over 4}
\ee
This metric has CTCs between $\r_- \simeq 0.246$ and $\r_+ \simeq 0.638$.
The behaviour of various quantities is illustrated in figure \ref{smoothex}. Note that there is no symmetry axis in this example, since $\sqrt{-g}$ is positive everywhere
and tends to zero for $\r \to - \infty$.
We see that the NEC is indeed violated in the ACTC region.
\FIGURE{
\begin{picture}(300,190)
\includegraphics[width=300pt]{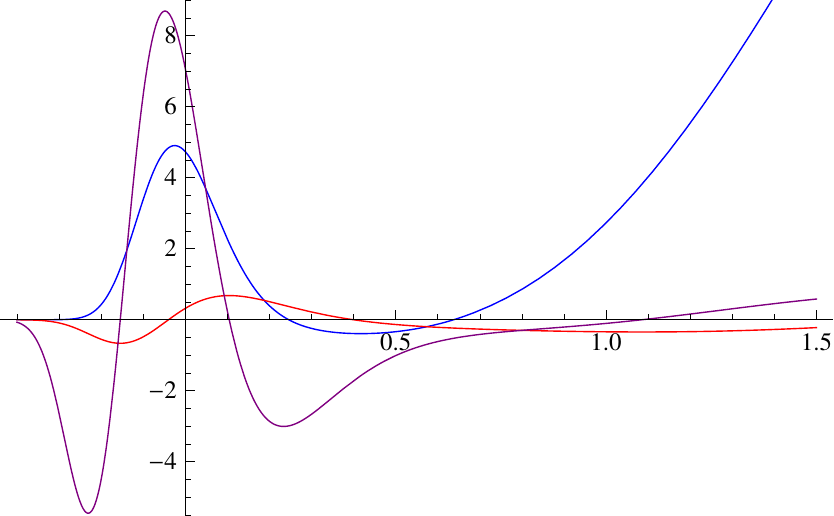}
\end{picture}
\caption{The smooth example with localized ACTCs discussed in the text. The blue curve is $g_{\f\f}$, the red one is $\sqrt{-g} v^\r_+$ and the purple one is $8\p G \sqrt{-g} T_{++}$ (the latter two are scaled down by a factor 15).}\label{smoothex}}

We can also quantitatively verify the main identity (\ref{mainid2}), in which we should take $\s = + 1$ since $l$ is positive everywhere. One finds that the radius $\r_c$ defined in (\ref{rhocdef}) is $\r_c \simeq 0.412$, and
\bea
 \int_{\r_-}^{\r_c} d\r \sqrt{-g}  T_{++}  &=& - {e^{2  B} \over 4 \sqrt{-g}} g_{tt} g_{\f\f}' (\r_- ) \simeq -5.595\\
 \int_{\r_c}^{\r_+} d\r \sqrt{-g}  T_{++}  &=& {e^{2  B} \over 4 \sqrt{-g}} g_{tt} g_{\f\f}' (\r_+ )\simeq -3.526
 \eea
 We can also verify  numerically the expression for $L_0$ in (\ref{enads}): the first term on the RHS, coming  from the ACTC region, contributes approximately -7.052 and the second term,
coming from outside  the ACTC region, contributes approximately 6.552, leading indeed to ${24 \over c} L_0= \half$.
We see that in this example, while the ACTC region contributes negatively to $L_0$, the total value of $L_0$ is still positive.

\section{Outlook} \label{concls}
In this paper we investigated   three-dimensional stationary, rotationally symmetric
spacetimes with CTCs. We found that these fall into two categories: they either have bad asymptotics (with CTCs present
all the way to infinity) or a matter stress tensor that violates the NEC by an amount given in (\ref{mainid}). In the asymptotically AdS case,
the NEC violating region leads to a negative contribution to the total value of $L_0$ according to  (\ref{enads}).
Let us conclude by pointing out some issues that merit further investigation.

In the context of classical general relativity, it would be useful to  extend our argument for violation of the NEC in CTC regions to less symmetric and higher dimensional spacetimes.
The assumption of rotational invariance was presumably not essential, since in its absence a triad with the required properties can still be found
as long as we can choose a Riemann normal coordinate $\r$ throughout the whole CTC region.

It would also be extremely interesting to bring the AdS/CFT correspondence to bear on the issue of chronology protection and pinpoint the pathologies
of spacetimes with CTCs. A first open question is whether CTCs in the bulk are always linked to violation of unitarity in the dual CFT.
In some examples, such as the G\"odel dust ball of \cite{Raeymaekers:2009ij}  reviewed in section \ref{dustball}, this is obviously the case:
 the violation of the NEC is so severe that the total value of $L_0$ is negative. Higher dimensional examples where CTCs in the bulk
 imply unitarity violation in the CFT  were found in \cite{Herdeiro:2000ap}.
In other examples, such as the one discussed in section \ref{smoothexsection}, there are CTCs in the bulk while $L_0$ remains positive.
This does not guarantee however that  unitarity is respected.
In such examples, on has to turn on a variety of terms in the metric which are  subleading  at large $\r$
and have the effect of driving $g_{\f\f}$ negative at finite values of $\r$.
These subleading terms encode one-point functions of
 other operators in the CFT than the stress tensor\footnote{For example, in the case of 3D gravity coupled to higher spins \cite{Campoleoni:2010zq}, these subleading terms encode the one-point functions of  primaries of conformal weight greater than two in the dual CFT.}, and are also expected to be constrained by unitarity. It would be interesting to explore these constraints further.  Another route to uncovering nonunitarity
 in the dual CFT could be the study of two-point functions. These are dominated by geodesics in the bulk connecting the two points on the boundary where the operators are inserted
 \cite{Balasubramanian:1999zv}. One would
 expect that, when the geodesic probes deep enough into the bulk (i.e. in the UV regime of the CFT), the correlator will receive contributions from geodesics looping
 around an arbitrary number of times in the CTC region, leading to pathologies.

 Since CTCs imply a violation of the NEC, the most attractive way to to rule out  CTCs would be to establish a relation between the NEC in the bulk and physical properties of  the boundary CFT.
It seems promising in this regard that the NEC  has been shown to be related to the c-theorem in the dual CFT \cite{Freedman:1999gp}.

\section*{Acknowledgements}
It is a pleasure to thank Shiraz Minwalla for the stimulating discussions that initiated this work, and Alena Pravdova to kindly explain various aspects of GR.
I have also benefitted from Matthew Headrick's mathematica package for tensor computations.
This work has been supported  in part by the Czech Science Foundation  grant GACR
P203/11/1388 and in part by the EURYI grant GACR  EYI/07/E010 from EUROHORC and ESF. I would also like to thank the Tata Institute of Fundamental Research, where
this work was initiated, for hospitality.

\end{document}